\documentclass[twoside]{article}

\usepackage{lipsum} % Package to generate dummy text throughout this template

\usepackage[sc]{mathpazo} % Use the Palatino font
\usepackage[T1]{fontenc} % Use 8-bit encoding that has 256 glyphs
\usepackage{microtype} % Slightly tweak font spacing for aesthetics
\usepackage{graphicx}
\usepackage[hmarginratio=1:1,top=32mm,columnsep=20pt]{geometry} % Document margins
\usepackage{multicol} % Used for the two-column layout of the document
\usepackage{hyperref} % For hyperlinks in the PDF

\usepackage[hang, small,labelfont=bf,up,textfont=it,up]{caption} % Custom captions under/above floats in tables or figures
\usepackage{booktabs} % Horizontal rules in tables
\usepackage{float} % Required for tables and figures in the multi-column environment - they need to be placed in specific locations with the [H] (e.g. \begin{table}[H])

\usepackage{lettrine} % The lettrine is the first enlarged letter at the beginning of the text
\usepackage{paralist} % Used for the compactitem environment which makes bullet points with less space between them

\usepackage{abstract} % Allows abstract customization
\usepackage{titlesec} % Allows customization of titles
\renewcommand\thesection{\Roman{section}}
\titleformat{\section}[block]{\large\scshape\centering}{\thesection.}{1em}{} % Change the look of the section titles

\usepackage{fancyhdr} % Headers and footers
\title{\vspace{-15mm}\fontsize{24pt}{10pt}\selectfont\textbf{Investigating the signatures of long-range persistence in seismic sequences along Circum-Pacific subduction zones}} % Article title

\author{
\large
\textsc{Daniel B. de Freitas}\\[2mm] % Your name
\normalsize Departamento de F\'{\i}sica, Universidade Federal do Cear\'a\\
\normalsize Caixa Postal 6030, Campus do Pici, 60455-900 Fortaleza, Cear\'a, Brazil \\ % Your institution
\normalsize \href{mailto:danielbrito@fisica.ufc.br}{danielbrito@fisica.ufc.br} % Your email address
\vspace{5mm}\\
\large
\textsc{G. S. Fran\c{c}a}\\[2mm] % Your name
\normalsize Observat\'orio Sismol\'ogico-IG/UnB, Campus Universit\'ario \\
\normalsize Darcy Ribeiro SG 13 Asa Norte, 70910-900 Bras\'{\i}lia, Brazil\\ % Your institution
\vspace{5mm}\\
\large
\textsc{T. M. Scheerer}\\[2mm] % Your name
\normalsize Conselho Nacional de Desenvolvimento Cient\'ifico e Tecnol\'ogico, CNPq, Brazil\\
\vspace{5mm}\\
\large
\textsc{C. S. Vilar}\\[2mm] % Your name
\normalsize Instituto de F\'{\i}sica, Universidade Federal da Bahia\\
\normalsize Campus Universit\'ario de Ondina, 40210-340 Salvador, Brazil\\ 
\vspace{5mm}\\
\large
\textsc{R. Silva}\\[2mm] % Your name
\normalsize Departamento de F\'{\i}sica,
Universidade Federal do Rio
Grande do Norte\\
\normalsize 59072-970, Natal,  RN, Brazil\\ 
\vspace{-5mm}\\
}
\date{}

\begin{document}

\maketitle % Insert title

\thispagestyle{fancy} % All pages have headers and footers

%----------------------------------------------------------------------------------------
%	ABSTRACT
%----------------------------------------------------------------------------------------

\begin{abstract}
	In the present paper, we analyze the signatures of long-range persistence in seismic sequences along Circum-Pacific subduction zones, from Chile to Kermadec, extracted from the National Earthquake Information Center (NEIC) catalog. This region, known as the Pacific Ring of Fire, is the world's most active fault line, containing about 90$\%$ of the world's earthquakes. We used the classical rescaled range ($R/S$) analysis to estimate the long-term persistence signals derived from a scaling parameter called the Hurst exponent, $H$. We measured the referred exponent and obtained values of $H>0.5$, indicating that a long-term memory effect exists. We found a possible fractal relationship between $H$ and the $b_{s}(q)$-index, which emerges from the non-extensive Gutenberg-Richter law as a function of the asperity. Therefore, $H$ can be associated with a mechanism that controls the level of seismic activity. Finally, we concluded that the dynamics associated with fragment-asperity interactions can be classified as a self-affine fractal phenomenon.
	
\end{abstract}

%----------------------------------------------------------------------------------------
%	ARTICLE CONTENTS
%----------------------------------------------------------------------------------------

\begin{multicols}{2} % Two-column layout throughout the main article text
	
	\section{Introduction}
	Geophysical signals often fluctuate in an irregular and complex way over time, and present inhomogeneous variations and extreme events, such as irregular rupture propagation and non-uniform distributions of rupture velocity, stress drop, and coseismic slip \cite{telesca0}. The presence of scaling properties in geophysical data points out that methods of fractal analysis based on the long-term correlations may provide a viable way to investigate the pattern of magnitudes in an episode of seismicity \cite{li}. The Earth's tectonic activity is due to very complex mechanisms that involve many variables, such as deformation, rupture, released energy, land features, and heterogeneity on the seismogenic plate interface \cite{kawa,scherrer}. To analyze such a complex time series, there is a plethora of different methods and tools that can be used to better describe the dynamical properties of earthquakes \cite{omori,gr}.
	
	Several statistical methods are reported in the scientific literature, which use the concept of fractality. Among them, we can find methods based on self-similar and self-affine fractals, such as the box dimension \cite{peitgen2004chaos}, the detrended fluctuation analysis (DFA) \cite{1992Natur.356..168P}, the detrending moving average analysis (DMA) \cite{ale}, and the scaled windowed variance analysis (SWVA) \cite{1985PhyS...32..257M}. 
	
	In the present paper, we will characterize the dynamics of earthquakes by calculating the Hurst exponent, where features such as long-range persistence can be investigated \cite{seuront}. Time series are quantified by their persistence or anti-persistence signature. Generally speaking, persistence can be described in terms of range, short- and long-range, where the memory can be classified as weak and strong, respectively. Statistical analysis is characterized by power-law distributions and can be a powerful tool for examining the temporal fluctuations at different scales when applied to earthquake magnitude time series \cite{telesca2}.
	
	In the present paper, we investigate the the signatures of long-range persistence that are present in the earthquake magnitude times series for the Circum-Pacific subduction zones, and have been already been processed by Scheerer \textit{et al}. \cite{scherrer}. Our study applied the rescaled range ($R/S$) analysis as a self-affine fractal method to the magnitude time series \cite{defreitas2013}. The characteristic measure of the $R/S$ analysis is the Hurst exponent, denoted by $H$. It is worth noting the universal character of the $R/S$ method in the analysis of the behavior of fluctuations. Many studies in different subject areas (e.g., economy, neuroscience, and astrophysics) have shown that the so-called Hurst exponent extracted from the $R/S$ analysis provides a robust and powerful statistical method to characterize nonstationary fluctuations at different timescales \cite{defreitas2013}, \cite{suyal} and \cite{li}. More recently, \cite{defreitas2013a} found a Hurst exponent of 0.87 for the San Andreas fault, which indicated a strong long-term persistence. Other studies (e.g., \cite{li}) also point out that the Hurst exponent is greater than 0.5, indicating a persistent behavior.
	
	Our main interest is to investigate a possible correlation between scaling properties (controlled by interaction asperities) \cite{lay} and the Hurst exponent estimated from the ($R/S$) analysis \cite{hurst1951,mw1969a}. In this paper, we show a detailed investigation of the rescaled range analysis to search for long-range correlations using the Hurst exponent in subduction zones along the Pacific Ring of Fire, the most seismically active region on Earth \cite{scherrer}. In particular, each zone was described using a general classification defined by an asperity model described by Scheerer \textit{et al}. \cite{scherrer}. Our main aim is to investigate two questions. Firstly, is there a correlation between $H$-value and the Circum-Pacific subduction zones? Secondly, is there any connection between the fractal parameter and the asperity model from Lay and Kanamori \cite{lay}? 
	
	Our paper is organized as follows, in the next section, we describe the Hurst method used in our study and a brief discussion about the nonextensive formalism. In section 3, we present a sample of our seismic catalog. The main results and their physical implications are presented in section 4. Finally, conclusions are drawn in Section 5.

	\begin{figure*}
		\includegraphics[width=0.80\textwidth,angle =270]{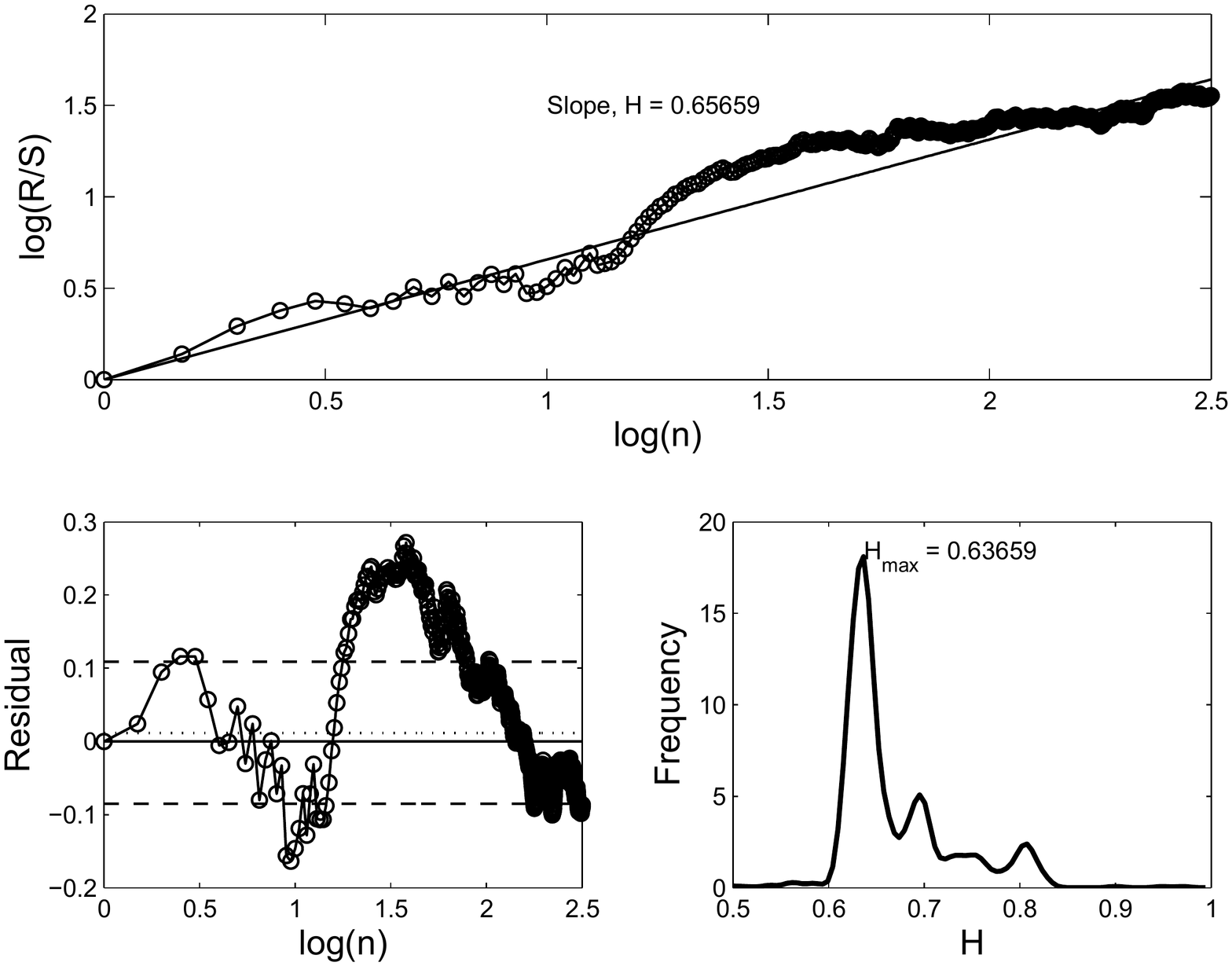}
		\caption{\textit{Top panel}: Individual $\log(R/S)$ points as a function of the logarithm of box-size $n$ for Alaska, representing the zone 1. \textit{Left bottom panel}: Residual extracted from the difference between the $\log(R/S)$ points and the best linear fit in the least-square sense. The solid line represents the perfect agreement, the dashed line denotes the mean value, whereas the dash-dotted lines indicate 1$\sigma$. \textit{Right bottom panel}: This plot is the Kernel adjustment of $H$ calculated as the derivative of the $R/S$ curve after each iteration $n$. $H_{max}$ is the maximum value of the distribution of $H$ which in all cases differs slightly from the value of $H$ identified by a straight line in the top panels.}
		\label{fig1}
	\end{figure*}

\begin{figure*}
	\includegraphics[width=0.80\textwidth,angle =270]{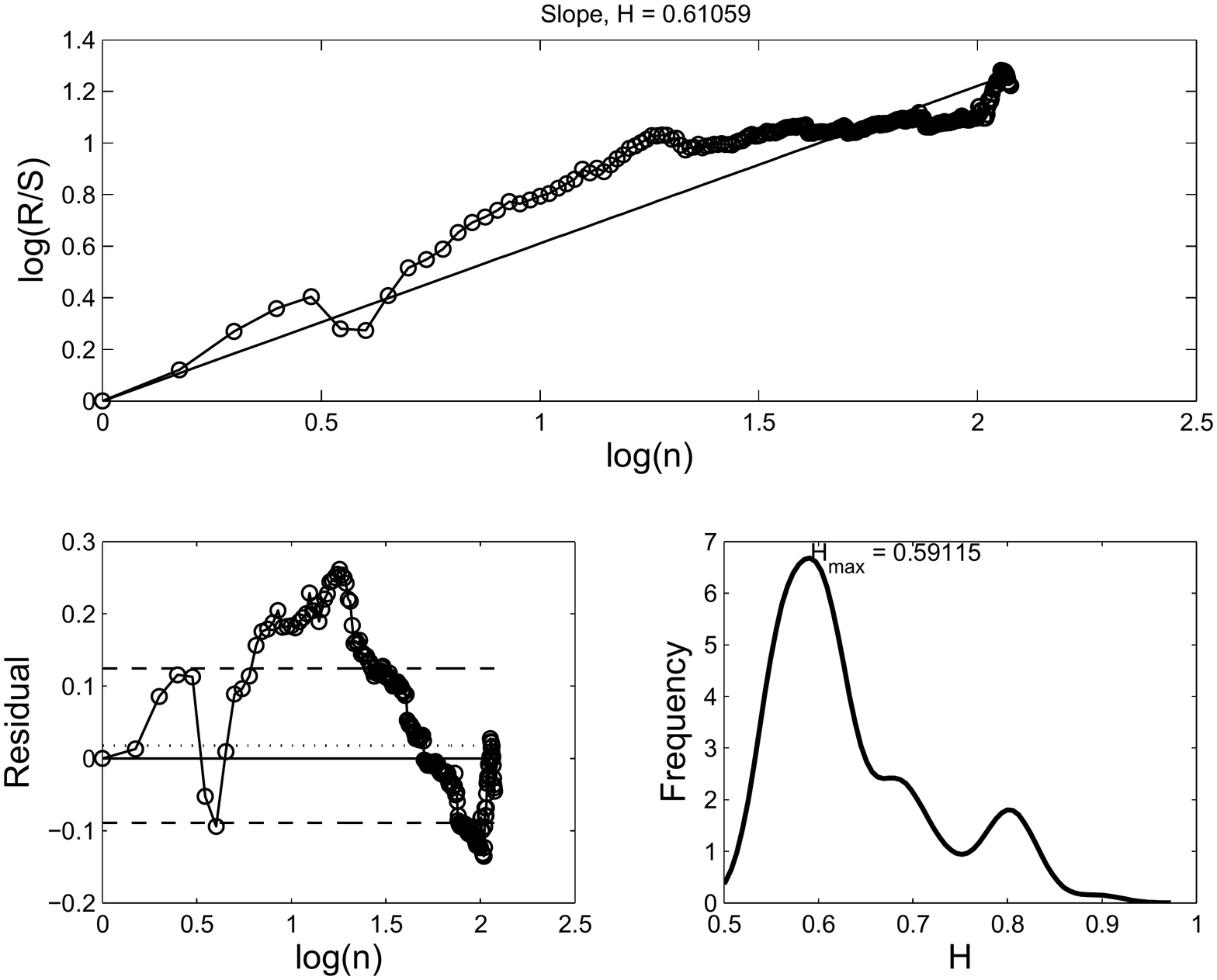}
	\caption{Fractal analysis for Colombia, representing the zone 2.}
	\label{fig1d}
\end{figure*}

\begin{figure*}
	\includegraphics[width=0.80\textwidth,angle =270]{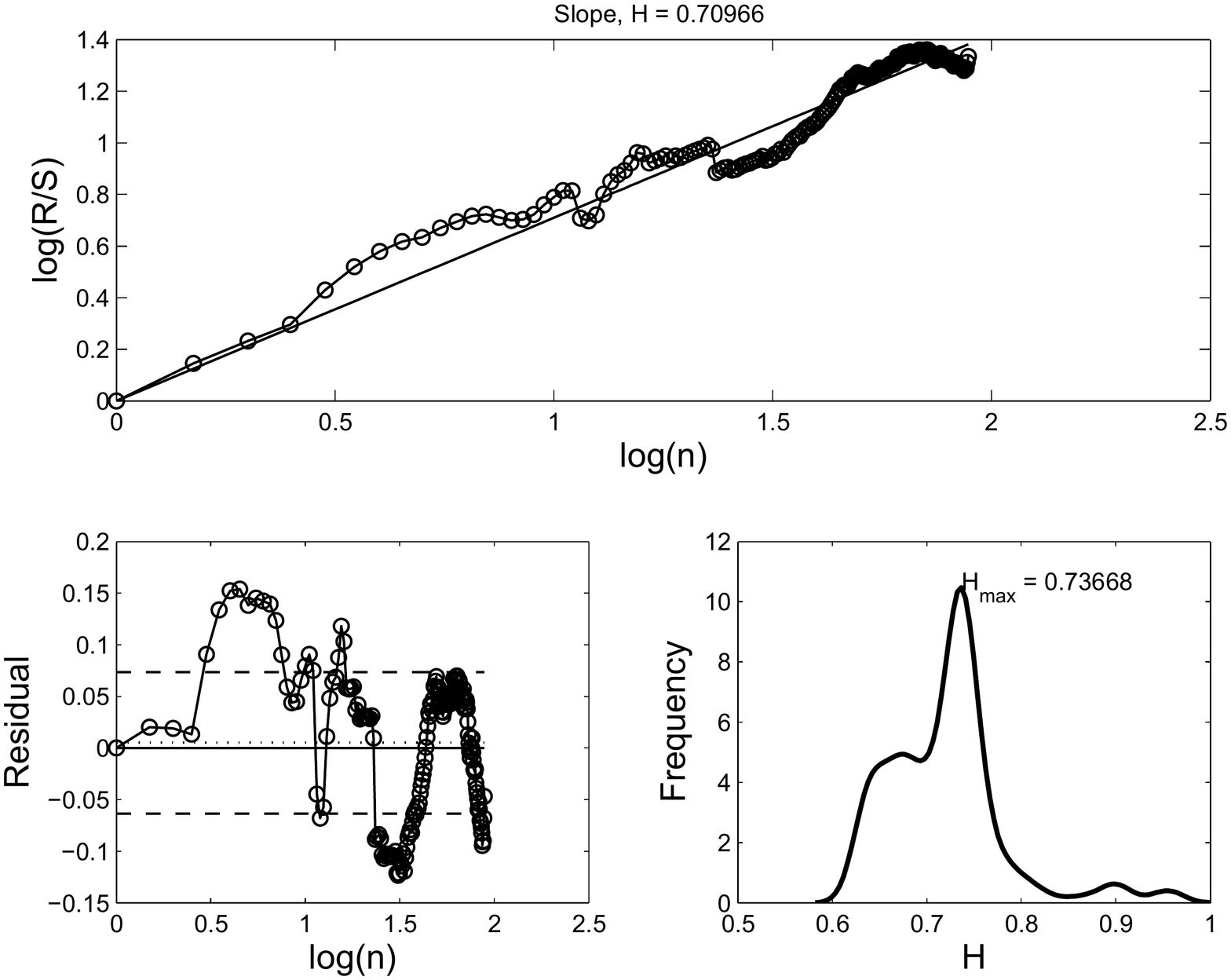}
	\caption{Fractal analysis for Peru, representing the zone 3.}
	\label{fig1b}
\end{figure*}

\begin{figure*}
	\includegraphics[width=0.80\textwidth,angle =270]{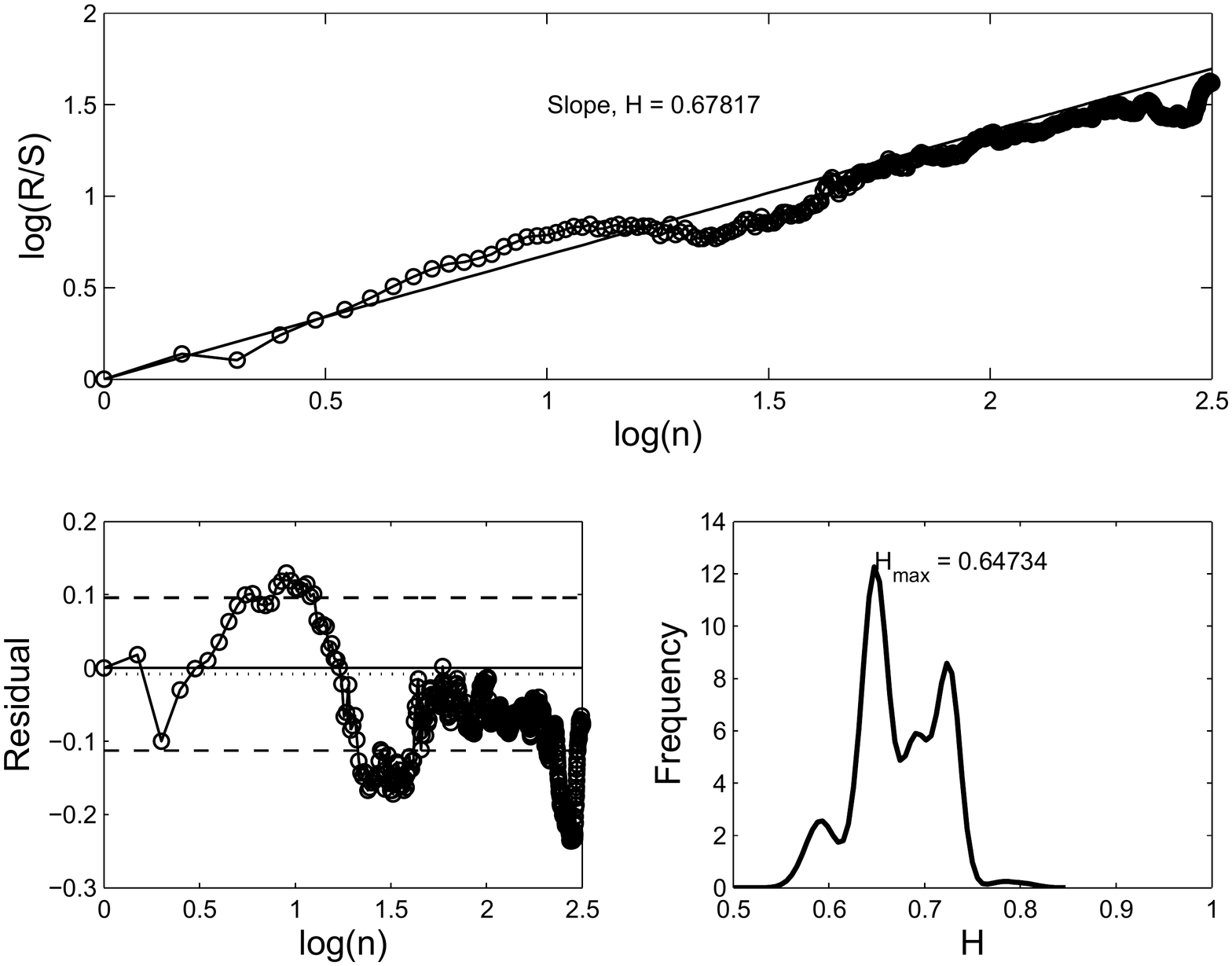}
	\caption{Fractal analysis for Marianas, representing the zone 4.}
	\label{fig1c}
\end{figure*}

	\begin{figure*}
		\includegraphics[width=0.80\textwidth,angle =270]{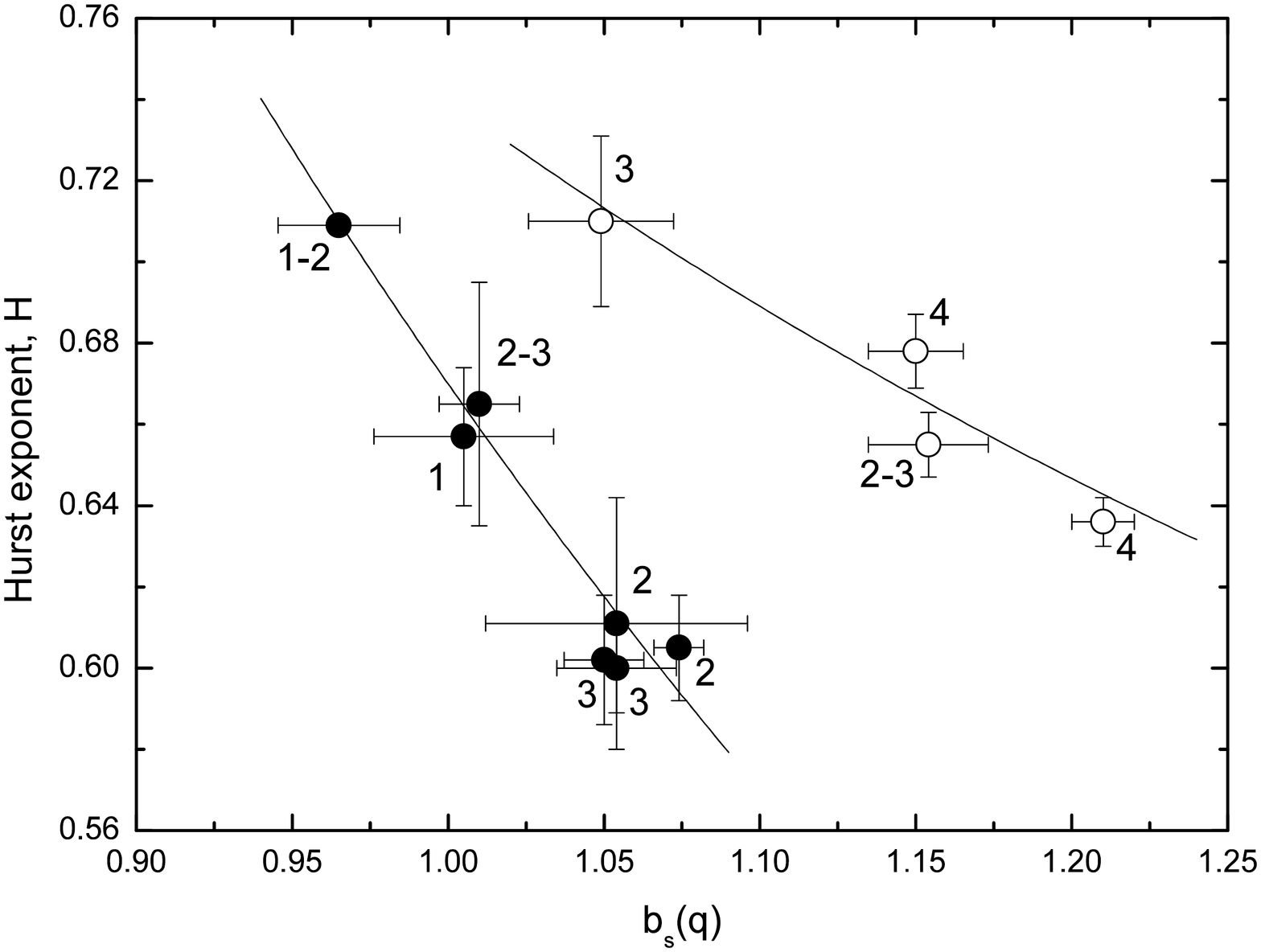}
		\caption{Values of the $b_{s}$-index extracted from the slope of the modified Gutenberg-Richter law as a function of $H$-index. Numbers indicate the category of each Subduction Zone. Solid and open circles are used to distinguish the two different domains. The numbers 1 to 4 denote 4 Subduction Zones indicated by Table 2.}
		\label{fig2}
	\end{figure*}
	
	\section{Statistical background}
	\subsection{Hurst effect}
	As pointed out by Seuront \cite{seuront}, there are various methods to describe the behavior of a time series using self-affine fractals. Mandelbrot and Wallis \cite{mw1969a} and \cite{mw1969b} introduced the concept of fractional Brownian motion (fBm) as a generalization of Brownian motion, which assumes the motion of an object is a union of rescaled copies of itself uniformly distributed in all directions, i.e., a self-similar fractal. In contrast, fBm considers the rescaling of the copies of itself is dependent on the direction, denoted as a self-affine \cite{seuront}. As mentioned by \cite{seuront}, a series of successive increments in a fBm defines a fractional Gaussian noise (fGn). This other type of time series based on the increment of a fBm yields a stationary signal with a mean of zero. Strictly speaking, a geophysical data set can be modeled as either a fGn-like or a noise-like time series \cite{defreitas2013a}.
	
	The scientific literature points out different techniques for exploring time series using a fractal approach. In general, the characteristics of a time series include a wider spectrum of complex measurements due to nonstationarity, nonlinearity, fractality, stochasticity, periodicity, chaos, and so on \cite{tang}. A powerful fractal technique for dealing with these assumptions is the Hurst analysis. As proposed by Hurst \cite{hurst1951}, we will focus on rescaled range analysis, also known as $R/S$ analysis.
	
	\subsubsection{Fractal index measured using the $R/S$ method}
	
As described by Hurst \cite{hurst1951}, a fractal analysis is used to measure the long-term memory or correlation of a time series. The method developed by Hurst \cite{hurst1951}, known as the rescaled range analysis (hereafter $R/S$ analysis), is the method adopted here to estimate the Hurst exponent $H$. Our aim is to verify the capability of the $R/S$ method to distinguish the different properties in the subduction zones through the behavior of $H$ and its possible correlation with the constant $b_{s}$, extracted from the generalized Gutenberg-Richter law, as measured by Sarlis and Skondas \cite{sarlis}. The aforementioned constant is the slope of the cumulative distribution number of earthquakes with magnitude greater than $m$ (for further details, see eq. 9 from \cite{sarlis}).
	
Firstly, we consider a time series given by $x(t): = x(1), x(2),..., x(N)$ with a time-window of length $N$. Following the same procedure mentioned by de Freitas \textit{et al}. \cite{defreitas2013}, the analysis starts with two elements, $n = 2$, and for each iteration one element is added until $n=N$ (i.e., the whole time series). In each cumulative window, we can measure two quantities that are denoted by $R(n)$ (the distance between the minimum and the maximum value of the accumulated deviations from the mean of $x(t)$ within the window of length $n$) and $S(n)$ (the standard deviation of the values of $x(t)$ in this same time window). From a mathematical point-view, $R(n)$ can be written as:
	\begin{equation}
		\label{eq1}
		R(n)=max_{1\leq t\leq n}X(t,n)-min_{1\leq t\leq n}X(t,n),
	\end{equation}
	where the variable $X(t,n)$ is defined as $X(t,n)=\sum^{n}_{t=1}\left(x(t)-\left\langle x(t)\right\rangle_{n}\right)$. Therefore, the long-trend is removed using the mean over the values of $x(t)$ within the window. $S(n)$ is defined as:
	\begin{equation}
		\label{eq2}
		S(n)=\left[\frac{1}{n}\sum^{n}_{t=1}\left(x(t)-\left\langle x(t)\right\rangle_{n}\right)^{2}\right]^{1/2}.
	\end{equation}

	Based on these two parameters, we define the $R/S$ statistics of the fluctuations in the time series as a power-law dependence over a box of $n$ elements given by:
	\begin{equation}
		\label{eq3}
		\frac{R(n)}{S(n)}=kn^{H},
	\end{equation}
	where $k$ is a constant and the Hurst index ($H$) can be measured by fitting the slope of the log-log plot of $R(n)/S(n)$ versus $n$: 
	\begin{equation}
		\label{eq3a}
		\log\left[\frac{R(n)}{S(n)}\right]\sim H\log(n).
	\end{equation}
	
	A time series described by the $R/S$ statistic is said to be fractal if the $R/S$ curve is a perfectly straight line; this means that the residual between this straight line and the $R/S$ curve must be null. If the $R/S$ curve presents some fluctuations (not large however) around the straight line that estimates $H$, this could be because the data are not ideal but observational and so are affected by measurement errors that make the curve $R/S$ depart (but not with large amplitude) from a straight line. If the $R/S$ curve does not follow a straight line or if the departures from the straight line are very large the object is not fractal.
	
	The Hurst exponent quantifies the probability that a given event in a process is followed by a similar event. As mentioned by de Freitas \textit{et al}. \cite{defreitas2013}, the $R/S$ method is used to calculate the scaling exponent, $H$, to give a quantitative measure of the persistence of a signal. Their typical values are: $0.5<H<1$ which indicates a persistence or long memory process, $H=0.5$ which indicates an uncorrelated process, and $0<H<0.5$ denotes anticorrelation. In a geophysical scenario, earthquakes bear dual features of randomicity and regularity, and therefore a $H$ value of between 0.5 and 1 is expected \cite{defreitas2013a}.
	
	\subsection{Nonextensive framework}
	Inspired by multifractals\footnote{Multifractals are a generalization of fractal systems in which only a single exponent (e.g., for instance, the Hurst exponent) is not enough to  describe their dynamics. Further details can be found in Refs. \cite{mw1969a,mw1969b,defreitas2017}.}, Tsallis \cite{tsallis1988} proposed a new concept of entropy extracted from a generalization of the Boltzmann-Gibbs (BG) entropy. This generalized entropy is defined as $S_{q}$ and is given by the following equation:
	\begin{equation}
		\label{eq8}
		S_{q}=-k_{B}\frac{1-\Sigma_{i=1}^{W}p^{q}_{i}}{q-1},
	\end{equation}
	which is based on the entropic index $q$ which measures the degree of nonextensivity of the system. In eq. (\ref{eq8}), $k_{B}$ denotes Boltzmann's constant, $W$ is the total number of microscopic states, and $p_{i}$ represents a set of probabilities. At the limit $q=1$, we recover the BG entropy. 
	
	By using this formalism, Sotolongo-Costa and Posadas \cite{costa04} and Silva \textit{et al}.\cite{silva} developed a new approach to describe the distribution of earthquakes with magnitude larger than $m$. According to Scherrer \textit{et al}. \cite{scherrer} and first demonstrated in \cite{sarlis}, the $b_{s}$ is related to entropic index $q$ by the expression
	\begin{equation}
		\label{eq5}
		b_{s}=\frac{2(2-q)}{q-1},
	\end{equation}
where $b_{s}$ is generally defined in the range between 0.8 and 1.2 (see Ref. \cite{sarlis} and references therein), and for the vast majority of systems studied so far, the $q$-index is limited to between 1 and 3 \cite{abe01}.
	
The Gutenberg-Richter law is an asymptotic relation between the total number of earthquakes $N$ and the magnitude $m$, given by:
	\begin{equation}
		\label{eq9}
		\log(N_{m})=a+b_{GR}m,
	\end{equation}
	where the $b_{s}$ values are calculated by Sarlis et al. \cite{sarlis} and the $b_{GR}$ values are calculated using the Gutenberg?Richter law cited above.
	
	As quoted by Scherrer \textit{et al}. \cite{scherrer}, the values of $b_{GR}$ are calculated using a software package to analyze the seismicity, denoted by ZMAP \cite{zmap}\footnote{\texttt{http://www.seismo.ethz.ch/en/research-and-teaching/products-software/software/ZMAP/}}. These authors found that the values of $b_{GR}$ differ from the $b_{s}$-index estimated using a nonextensive approach. They also found that the $q$-values correlated with some properties of subduction zones, such as the occurrence of ruptures, seismic/aseismic slip, coupling, and the interaction of asperities\cite{scherrer}.
	
	\section{Catalog data}
	Scherrer \textit{et al}. \cite{scherrer} have produced a list of four Circum-Pacific subduction zones distributed in a belt along the so-called Ring of Fire (see Fig.1 from the aforementioned paper). These data were extracted from the National Earthquake Information Center (NEIC) catalog\footnote{\texttt{https://www.usgs.gov/natural-hazards/earthquake-hazards/data-tools}}\cite{neic}. From that sample, we selected 12 areas to apply our analysis, with 142,280 events in the magnitude interval $1<m<9$ during a decade from 2001 to 2010 (see Table \ref{tab1}). A map with the distribution of the Circum-Pacific subduction zones can be seen in Figure 1 from Scherrer \textit{et al}. \cite{scherrer}. As reported by \cite{scherrer}, the NEIC catalog offers magnitude time series with different magnitudes types ($M_{w}, M_{b}, M_{s}, M_{l}, and M_{d}$) for the same event. In addition, we decide that using this sequence makes no significant impact on the result of the present paper because the differences between magnitudes types are smaller than 1. As the $M_{s}$ (surface wave magnitude) is rarely used, we did not consider the bias from this magnitude \cite{reza}.
	
	The data sample used by Scherrer \textit{et al}. \cite{scherrer} is distributed in four different subduction zones defined by asperities and broadness of the rupture front. The main structure of the zones is described by the authors in the aforementioned study. The reader is referred to Scherrer \textit{et al}. \cite{scherrer} for details regarding the instrumental procedure and classification.
	
	For the present analysis, we considered earthquakes with a magnitude greater than 1 for all the regions, in this case only the effect due to macroearthquakes is analyzed. Scherrer \textit {et al}. \cite{scherrer} measured three important parameters: (i) the entropic index $q$, which emerges from the nonextensive statistical mechanics \cite{tsallis1988}, \cite{abe01} and \cite{gell2004}, (ii) the $b_{s}$-index extracted from the slope of the nonextensive Gutenberg-Richter law \cite{costa04} and \cite{silva}, and (iii) the classical index $b_{GR}$ from the Gutenberg-Richter law \cite{gr} calculated using ZMAP software. In the next section, we compare the Gutenberg-Richter indexes with the Hurst exponent $H$\footnote{The values of $H$-index were calculated using MATLAB code indicated by the link: \texttt{https://www.mathworks.com/matlabcentral/fileexchange/39069-hurst-exponent-estimation}}.

	\begin{table*}
		\caption{Identifier number of subduction zones $(SZ)$ and $b_{s}$ (see Ref. \cite{scherrer}), and $H$ and $\sigma^{H}_{Kernel}$ estimated by our analysis. Symbols $\circ$ and $\bullet$ are used to differentiate the subsamples shown in Fig.\ref{fig2}.}
		\label{tab1}
		\begin{center}
			\begin{tabular}{lccccc}
				\hline
				Area & $SZ$ & $b_{S}\pm\sigma$  & $H\pm\sigma$ \\
				&  &  & & \\
				\hline
				Alaska $\bullet$ & 1&1.005$\pm$0.019 & 0.657$\pm$0.017\\
				Aleutians $\bullet$ & 1-2&0.965$\pm$0.029 & 0.709$\pm$0.002\\
				Central America $\circ$ & 2-3&1.154$\pm$0.013 & 0.655$\pm$0.008\\
				Central Chile $\bullet$ & 3 &1.054$\pm$0.023 & 0.600$\pm$0.011\\
				Colombia $\bullet$ & 2 &1.054$\pm$0.015 &0.611$\pm$0.031\\
				Kuriles $\bullet$ & 3 &1.050$\pm$0.042 & 0.602$\pm$0.016\\
				Marianas $\circ$ & 4 &1.150$\pm$0.015 & 0.678$\pm$0.009\\
				New Hebrides $\bullet$ & 2-3&1.010$\pm$0.008 & 0.665$\pm$0.03\\
				Peru $\circ$ &3 &1.049$\pm$0.015 & 0.710$\pm$0.02\\
				Solomon Islands $\bullet$ &2 &1.074$\pm$0.019 & 0.605$\pm$0.013\\
				Tonga \& Kermadec $\circ$ & 4&1.210$\pm$0.01 & 0.636$\pm$0.006\\
				\hline
			\end{tabular}
		\end{center}
	\end{table*}
	
	\section{Results and discussions}
	As shown in the top panels in Figures from \ref{fig1} to \ref{fig1d}, the $R/S$ method was used to estimate the values of the Hurst exponent for a data set of four Circum-Pacific subduction zones. As a result, the values of $H$ were calculated using the slope of the log-log plot of $R/S$ versus $n$, over the entire range of $n$ and are summarized in Table \ref{tab1}. 
	
In the right bottom panels in Figures \ref{fig1} to \ref{fig1d}, there is a spectrum of the $H$ exponent, which was calculated using a kernel density estimation. As the Kernel density is based on the smoothing functions, we used it to identify the profile of the distribution of the $H$-index, and therefore to estimate the width of the $H$ distribution. In certain cases, the width of the distribution of $H$ is quite narrow, and therefore the data sample can be considered a fractal. According to the residual panels, such as those shown in Figures \ref{fig1} to \ref{fig1d}, we found that there is a considerable variation in the value of $H$ along the magnitude series. However, we would have to use multifractal methods for greater precision regarding how narrow the distribution should be in order for the system to be considered a fractal. 
	
We performed a bootstrap resampling method to estimate the 95\% confidence interval in each time series. Firstly, we used a set of 1000 bootstrap replications of the Hurst exponent calculated point to point, i.e., for each ratio of $R(n)/S(n)$ one value of $H$ can be extracted so one time series yields $n-2$ values of $H$. In this case, we have a set of values for $H$, then we performed the bootstrap method. Finally, we ranked the bootstrapped means of $H$, from the lower to the higher value, and took the 25th and the 75th means in the rank as the lower and upper limits of the confidence interval, respectively. The mean Hurst exponents and their symmetrical confidence interval are presented in Table \ref{tab1}.
	
As seen in Table \ref{tab1} all the values of these exponents suggest that there is persistence in the subduction zone data, i.e., the values of $H$ are always greater than 0.5, mostly fluctuating around 0.65. In this sense, the values of $H>0.5$ might show that earthquakes in the Pacific Ring of Fire are not a Gaussian process, i.e., random variables are not normally distributed. On the contrary, there is a long-term memory associated with the fluctuation dynamics \cite{li}.
	
Firstly, we verified that only the $b_{s}$-index provides a reasonable correlation with $H$. This relationship can be observed in Fig. \ref{fig2}, where we divided the data points into two regimes with different slopes. We fit the following law to these regimes: 
	\begin{equation}
		\label{eq10}
		H=\frac{A}{b_{s}}+C.
	\end{equation}
We found the following values for slopes: $A_{\circ}=0.56\pm 0.11$ and $A_{\bullet}=1.10\pm 0.10$. The values for the intercepts were: $C_{\circ}=0.18\pm 0.12$ and $C_{\bullet}=-0.43\pm 0.11$. We calculated the anti-correlation using the Spearman ($r_{S}$) and Pearson ($r_{P}$) correlation coefficients and found that $r^{\circ}_{S}=-0.99$, $r^{\circ}_{P}=-0.96$, $r^{\bullet}_{S}=-0.74$, and $r^{\bullet}_{P}=-0.97$ \cite{press}. We observed that $A_{\bullet}$ was associated with subduction zones with large magnitudes, whereas $A_{\circ}$ was associated with areas with smaller magnitudes. In particular, the values of $H$ can be associated with the mechanism which controls the level of seismic activity.
	
	\subsection{Is it possible to determine the index $H$ from $b_{S}$?}
	
	Inspired by the works of Borland \cite{borland}, Sarlis \textit{et al}. \cite{sarlis} and Scherrer \textit{et al}. \cite{scherrer}, and by the empirical relation found in the previous section, we investigated a possible theoretical correlation between the indexes $H$ and $b_{s}$, based on the entropic index $q$. 
	
	Borland \cite{borland} introduced a relation between the $q$-index and the exponent $H$:
	\begin{equation}
		\label{eq6}
		H=\frac{1}{3-q}.
	\end{equation}
	The above relationship is obtained by using the Langevin equation as a function of the entropic index $q$ \cite{borland}. This relationship is only valid for $-\infty<q<2$ because the range of the $H$-index is defined between zero and unity.
	
	The equations (\ref{eq5}) and (\ref{eq6}) suggest that it is possible to determine a relation between $H$ and $b_{s}$. Substituting $q$ from (\ref{eq5}) into (\ref{eq6}), we have
	\begin{equation}
		\label{eq7}
		H=0.5+\frac{0.5}{1+b_{s}},
	\end{equation}
	which indicates an anti-correlation between the indexes.
	
	As we can observe in Figure \ref{fig2}, the above equation is not in agreement with the values found for the parameters $A$ and $C$, which were extracted from an empirical relationship (\ref{eq10}). Instead, equation (\ref{eq7}) can be used to estimate the lower and upper values of $H$. Based on the expected values of $b_{s}$ mentioned in Section II.2, the values of $H$ are limited to between 0.72 and 0.78. However, these theoretical values of $H$ are an overestimation when compared with the values shown in Table\ref{tab1}, indicating that equation (\ref{eq7}) cannot be used to estimate the values of $H$.

	\section{Final remarks}
	We aimed to answer the two questions previously mentioned in the introduction. Firstly, is there a correlation between the $H$-value and the Circum-Pacific subduction zones? Secondly, is there any connection between the fractal parameter and the asperity model from Lay and Kanamori \cite{lay}? Our study indicates the possible geophysical meaning of the $H$-index in relation to subduction zones. As seen in section 4, there are two different behaviors for the $H$-value as a function of the parameter $b_{s}$. The explication for these two domains of the value of $H$ may be the lack of large earthquakes during this time period (2001-2010) in the Peru- and Marianas-type zones. In fact, zones 3 and 4 present large amounts of aseismic slip, where there is an inhibition of large rupture development, generating complicated ruptures and foreshock?aftershock activity.
	
	We used the Hurst analysis to investigate the behavior of twelve magnitude catalogs along the Pacific Ring of Fire. From that analysis, we found that there is a relationship between the Hurst exponent $H$ and the modified Gutenberg-Richter index $b_{s}$, as illustrated by Fig. \ref{fig2}. 
	
	Our results reveal a strong long-range persistence in the magnitudes series studied, indicated by the values of $H$ greater than 0.5. The long-range persistence shows that the coupling between random variables in magnitude series at different times is stronger than the short-range one. In other words, the persistent behavior in the time series characterized by calculating the Hurst exponent reveals the existence of the persistence of long-range memory. In addition, there is a statistical correlation between the subduction zones, as shown in Figure \ref{fig2}.
	
The existence of an empirical correlation between the Hurst exponent and the $b_{s}$-index opens up the possibility of proposing a better model for seismic hazard and risk assessments with high memory, which these systems present. In addition, we can conclude that the dynamics associated with fragment-asperity interactions can be emphasized as a self-affine fractal phenomenon.

%----------------------------------------------------------------------------------------
%	REFERENCE LIST
%----------------------------------------------------------------------------------------

%----------------------------------------------------------------------------------------

\end{multicols}

\end{document}